\documentclass[twocolumn]{aastex63}

\received{}
\revised{}
\accepted{}
\submitjournal{ApJ}

\shorttitle{Carbon Chain vs. COM Chemistry in Low-Mass YSOs in the Perseus Region}
\shortauthors{Taniguchi et al.}

\begin{document}

\title{Carbon-Chain Chemistry vs. Complex-Organic-Molecule Chemistry in Envelopes around Three Low-Mass Young Stellar Objects in the Perseus Region}

\correspondingauthor{Kotomi Taniguchi, Liton Majumdar}
\email{kotomi.taniguchi@gakushuin.ac.jp; liton@niser.ac.in}

\author[0000-0003-4402-6475]{Kotomi Taniguchi}
\affiliation{Department of Physics, Faculty of Science, Gakushuin University, Mejiro, Toshima, Tokyo 171-8588, Japan}

\author[0000-0001-7031-8039]{Liton Majumdar}
\affiliation{School of Earth and Planetary Sciences, National Institute of Science Education and Research, HBNI, Jatni 752050, Odisha, India}

\author[0000-0003-0845-128X]{Shigehisa Takakuwa}
\affiliation{Department of Physics and Astronomy, Graduate School of Science and Engineering, Kagoshima University, Korimoto, Kagoshima, Kagoshima 890-0065, Japan}

\author[0000-0003-0769-8627]{Masao Saito}
\affiliation{National Astronomical Observatory of Japan (NAOJ), National Institutes of Natural Sciences, Osawa, Mitaka, Tokyo 181-8588, Japan}
\affiliation{Department of Astronomical Science, School of Physical Science, SOKENDAI (The Graduate University for Advanced Studies), Osawa, Mitaka, Tokyo 181-8588, Japan}

\author[0000-0002-0500-4700]{Dariusz C. Lis}
\affiliation{Jet Propulsion Laboratory, California Institute of Technology, 48010 Oak Grove Drive, Pasadena, CA 91109, USA}

\author[0000-0002-6622-8396]{Paul F. Goldsmith}
\affiliation{Jet Propulsion Laboratory, California Institute of Technology, 48010 Oak Grove Drive, Pasadena, CA 91109, USA}

\author[0000-0002-4649-2536]{Eric Herbst}
\affiliation{Department of Astronomy, University of Virginia, Charlottesville, VA 22904, USA}
\affiliation{Department of Chemistry, University of Virginia, Charlottesville, VA 22903, USA}



\begin{abstract}
We have analyzed ALMA Cycle 5 data in Band 4 toward three low-mass young stellar objects (YSOs), IRAS 03235+3004 (hereafter IRAS 03235), IRAS 03245+3002 (IRAS 03245), and IRAS 03271+3013 (IRAS 03271), in the Perseus region.
The HC$_{3}$N ($J=16-15$; $E_{\rm {up}}/k = 59.4$ K) line has been detected in all of the target sources, while four CH$_{3}$OH lines ($E_{\rm {up}}/k = 15.4-36.3$ K) have been detected only in IRAS 03245.
Sizes of the HC$_{3}$N distributions ($\sim 2930-3230$ au) in IRAS 03235 and IRAS 03245 are similar to those of the carbon-chain species in the warm carbon chain chemistry (WCCC) source L1527.
The size of the CH$_{3}$OH emission in IRAS 03245 is $\sim 1760$ au, which is slightly smaller than that of HC$_{3}$N in this source.
We compare the CH$_{3}$OH/HC$_{3}$N abundance ratios observed in these sources with predictions of chemical models.
We confirm that the observed ratio in IRAS 03245 agrees with the modeled values at temperatures around 30--35 K, which supports the HC$_{3}$N formation by the WCCC mechanism.
In this temperature range, CH$_{3}$OH does not thermally desorb from dust grains. 
Non-thermal desorption mechanisms or gas-phase formation of CH$_{3}$OH seem to work efficiently around IRAS 03245.
The fact that IRAS 03245 has the highest bolometric luminosity among the target sources seems to support these mechanisms, in particular the non-thermal desorption mechanisms.
\end{abstract}



\section{Introduction} \label{sec:intro}

Studying the environment and evolution of low-mass star forming regions is important for revealing how our Sun was born.
The chemical composition in star-forming regions provides us various information, for example, evolutionary stages and physical conditions \citep[e.g.,][]{2012A&ARv..20...56C}.

The gas-phase chemical composition around young stellar objects (YSOs) yields essential information on not just the current evolutionary stage but also former environments or evolutionary processes in the pre-stellar core phase \citep{2013ChRv..113.8981S, 2016A&A...592L..11S, 2019ApJ...881...57T, 2020MNRAS.493.2395T}.
The ice mantles form in cold and dense starless cores.
Although saturated complex organic molecules (COMs) were considered to be deficient in the gas phase during the cold stage, several COMs have now been detected in starless cores, e.g., the Barnard 5 cold dark cloud \citep{2017A&A...607A..20T} and the L1544 pre-stellar core \citep{2016ApJ...830L...6J}.
These COMs are considered to be formed in the gas phase \citep[e.g.,][]{2015MNRAS.449L..16B} or formed on dust surfaces followed by non-thermal desorption mechanisms.
New dust-surface mechanisms proposed by \citet{2020ApJS..249...26J} successfully reproduce observed abundances of COMs in the L1544 pre-stellar core.
Furthermore, modeling studies including radiolysis can also produce high COM abundances \citep[e.g.,][]{2018PCCP...20.5359S, 2018ApJ...861...20S, 2021MNRAS.500.3414P}.
Hence, even in cold dense cores, chemical processes forming COMs are also efficient.
After YSOs are born, ice mantles, including COMs, sublimate when the temperature rises to $\approx 100$ K.
The chemistry in YSOs then strongly depends on their density and temperature structures \citep[e.g.,][and therein]{2020ARA&A..58..727J}.

There are two, well-recognized chemical processes effective around low-mass YSOs.
One is hot corino chemistry, in which saturated COMs are abundant in the hot ($>100$ K) dense ($\geq10^{7}$ cm$^{-3}$) gas \citep{2004ASPC..323..195C, 2009ARA&A..47..427H}.
Hot corino chemistry is initiated by successive hydrogenation reactions of CO molecules on grain surfaces to form CH$_{3}$OH, which is a fundamental COM and a precursor of more complex COMs \citep{2020arXiv201003529O}.
The other is warm carbon chain chemistry \citep[WCCC;][]{2013ChRv..113.8981S}.
Carbon-chain molecules are formed from gaseous CH$_{4}$, which is one of the main constituents of ice mantles \citep{2020arXiv201003529O}, in lukewarm envelopes \citep[$\approx 25-35$ K,][]{2008ApJ...681.1385H}.
Hence, the presence of two types of chemistry around low-mass YSOs may indicate that some conditions of the prestellar core phase produce CO-rich ice and others lead to CH$_{4}$-rich ice.
However, the origin of the chemical diversity is still controversial. 
The results of this paper will help to explain the chemical diversity.

Although COMs are generally considered to be abundant in hot dense gas around YSOs, a recent survey observation has revealed that CH$_{3}$OH and CH$_{3}$CHO are prevalent.
\citet{2020ApJ...891...73S} conducted survey observations of these COMs toward 31 starless cores in the Taurus region with the Arizona Radio Observatory (ARO) 12 m telescope, and reported that the detection fractions of CH$_{3}$OH and CH$_{3}$CHO are 100\% (31/31) and 70\% (22/31), respectively.
It was also found that the CH$_{3}$OH abundance decreases from young cores to evolved starless cores possibly due to depletion onto dust grains \citep{2020ApJ...891...73S}.
The detection of COMs in the starless cores suggests that there are some non-thermal desorption mechanisms or efficient gas-phase formation pathways at work in starless cores.

\citet{1998ApJ...501..723T} showed different spatial distributions of H$^{13}$CO$^{+}$ and CH$_{3}$OH in the TMC-1C region and suggested that CH$_{3}$OH cores are younger and farther from protostar formation than H$^{13}$CO$^{+}$ cores.
A subsequent study supports the possibility of the different evolutionary stages of these cores \citep{2003ApJ...584..818T}.
Furthermore, \citet{2000ApJ...542..367T} carried out mapping observations toward the Heiles Cloud 2 region in Taurus and found that the CH$_{3}$OH emission is weak toward the protostars (Class 0 and Class I) and rather stronger toward cores without protostars.
\cite{2003ApJ...590..932T} also conducted mapping observations toward IRAS 04191+1522, a very young Class 0 YSO.
They found that the CH$_{3}$OH abundance toward the YSO is $\sim 10$ times lower than that toward the interaction region between the outflow and the surrounding dense gas. 
Similar observational results of the non-detection or weak emission of CH$_{3}$OH at protostellar positions have been reported \citep[e.g.,][]{2006ApJ...651L..41W}.
The weak CH$_{3}$OH emission at protostellar cores has been reported even in the higher angular-resolution observations with ALMA at 3 mm, where the obscuring effect by the dust emission is not significant \citep{2020MNRAS.499.4394M}.
All these observational results imply that CH$_{3}$OH is associated with young starless cores and adsorbed onto dust grains in evolved starless cores.
The weak CH$_{3}$OH emission toward low-mass YSOs \citep{2000ApJ...542..367T, 2003ApJ...590..932T, 2006ApJ...651L..41W, 2020MNRAS.499.4394M} probably indicates that CH$_{3}$OH lines are not necessarily tracers for low-mass protostellar cores, and not all the protostellar sources possess hot core/hot corinos.

Cyanoacetylene (HC$_{3}$N), which is the shortest member of the cyanopolyyne family (HC$_{2n+1}$N), is known to be abundant in starless cores, as well as other carbon-chain species \citep[e.g.,][]{1992ApJ...392..551S}.
It is also detected around low-mass and high-mass YSOs \citep{2018ApJ...863...88L, 2018ApJ...854..133T,  2019ApJ...872..154T}, and protoplanetary disks \citep{2015Natur.520..198O, 2018ApJ...857...69B}.
Hence, HC$_{3}$N is prevalent in star-forming regions in various evolutionary stages.
However, it has not been determined whether the WCCC mechanism is responsible for formation of HC$_{3}$N around low-mass YSOs, because there is no information about spatial distributions of HC$_{3}$N around these sources at sufficiently high spatial resolutions  \citep{2018ApJ...863...88L}.

In the present paper, we focus on methanol and cyanoacetylene as potential indicators to constrain evolutionary stages of low-mass YSOs. 
For this purpose, high angular-resolution imaging observations of CH$_{3}$OH and HC$_{3}$N in low-mass YSOs are essential. 
We investigate the spatial distributions of HC$_{3}$N and CH$_{3}$OH toward three low-mass YSOs, IRAS 03235+3004 (hereafter IRAS 03235), IRAS 03245+3002 (IRAS 03245), and IRAS 03271+3013 (IRAS 03271), in the Perseus region, using ALMA archival data. 
In Section \ref{sec:data}, we will describe the used data and reduction procedures.
The resultant continuum images, moment 0 maps and spectra of HC$_{3}$N and CH$_{3}$OH are shown in Section \ref{sec:resana}.
The observed CH$_{3}$OH/HC$_{3}$N abundance ratio is compared with the results of the chemical network simulations in Section \ref{sec:dis}.
Finally, we summarize our main conclusions in Section \ref{sec:con}.

\section{Data Reduction} \label{sec:data}

In this paper, we present ALMA Band 4 archival data toward three low-mass YSOs in the Perseus region taken from Cycle 5 data\footnote{project ID; 2017.1.00955.S, PI; Jennifer Bergner}.
The properties of our target sources are summarized in Table \ref{tab:source}.
In Table \ref{tab:source}, the coordinates correspond to the phase reference centers of these ALMA observations carried out with the 7-m Array in 2018 July.

Table \ref{tab:spw} summarizes information concerning each spectral window presented in this paper.
We used the first spectral window for continuum data.
The second and third windows contain lines of HC$_{3}$N and CH$_{3}$OH, respectively.
Observed lines of each molecule are summarized in Table \ref{tab:line}.
The frequency resolution of 121 kHz corresponds to a velocity resolution of $\sim 0.24$ km s$^{-1}$.
The field of view (FoV) and largest angular scale (LAS) are 65\farcs3 and 43\farcs6, respectively.

\begin{deluxetable*}{ccccccccc}
\tablenum{1}
\tablecaption{Summary of target sources \label{tab:source}}
\tablewidth{0pt}
\tablehead{
\colhead{Source Name} & \colhead{R.A. (J2000)\tablenotemark{a}} & \colhead{Decl. (J2000)\tablenotemark{a}} & \colhead{$T_{\rm {bol}}$ (K)\tablenotemark{b}} & \colhead{$L_{\rm {bol}}$ (L$_{\sun}$)\tablenotemark{c}} & \colhead{$M_{\rm {env}}$ (M$_{\sun}$)\tablenotemark{d}} & \colhead{$\alpha$\tablenotemark{b}} & \colhead{Class\tablenotemark{e}} & \colhead{$T_{\rm {dust}}$ (K)\tablenotemark{f}}
}
\startdata
IRAS 03235+3004 & $03^{\rm {h}}26^{\rm {m}}$37\fs45 & +30\degr15\arcmin27\farcs9 & 73 & 1.29 & $0.69 \pm 0.11$ & 1.03 & 0  & ... \\
IRAS 03245+3002 & $03^{\rm {h}}27^{\rm {m}}$39\fs03 & +30\degr12\arcmin59\farcs3 & 65 & 4.94 & $0.70 \pm 0.04$ & 2.46 & 0 & $37 \pm 2$ \\
IRAS 03271+3013 & $03^{\rm {h}}30^{\rm {m}}$15\fs16 & +30\degr23\arcmin48\farcs8 & 100 & 2.06 & $0.63 \pm 0.10$ & 1.58 & I & $45 \pm 2$ \\
\enddata
\tablenotetext{a}{Coordinates of the phase reference centers.}
\tablenotetext{b}{Bolometric temperature taken from \citet{2015ApJS..220...11D}.}
\tablenotetext{c}{Bolometric luminosities at an assumed distance of 250 pc taken from \citet{2015ApJS..220...11D} are scaled to the newly measured distance of Perseus \citep[293 pc;][]{2018ApJ...865...73O}.}
\tablenotetext{d}{Envelope masses at an assumed distance of 250 pc taken from \citet{2009ApJ...692..973E} are scaled to the newly measured distance of Perseus (293 pc).}
\tablenotetext{e}{IR spectral indexes ($\alpha$) taken from \citet{2009ApJ...692..973E}.}
\tablenotetext{f}{Dust temperatures taken from \citet{2009AA...493...89E}.}
\end{deluxetable*}

\begin{deluxetable}{cccc}
\tablenum{2}
\tablecaption{Summary of spectral windows covered by the correlator setup \label{tab:spw}}
\tablewidth{0pt}
\tablehead{
\colhead{Frequency} & \colhead{Frequency} & \colhead{Angular} & \colhead{Molecular} \\
\colhead{range (GHz)} & \colhead{res. (kHz)} & \colhead{resolution} & \colhead{Lines}
}
\startdata
147.097--147.224 & 124 & 11\farcs0 $\times$ 8\farcs6 & Continuum \\
145.526--145.591 & 121 & 11\farcs2 $\times$ 8\farcs4 & HC$_{3}$N \\
157.236--157.301 & 121 & 10\farcs6 $\times$ 8\farcs2 & CH$_{3}$OH \\
\enddata
\end{deluxetable}

\begin{deluxetable}{clccc}
\tablenum{3}
\tablecaption{Summary of target molecular lines \label{tab:line}}
\tablewidth{0pt}
\tablehead{
\colhead{Molecule} & \colhead{Transition} & \colhead{Frequency} & \colhead{$E_{\rm {up}}/k$} & \colhead{log$_{10}$($A_{\rm {ij}}$)} \\
\colhead{} & \colhead{} & \colhead{(GHz)} & \colhead{(K)} & \colhead{(s$^{-1}$)}
}
\startdata
HC$_{3}$N & 16--15 & 145.560946 & 59.4 & -3.61727 \\
CH$_{3}$OH & $4_{0}-4_{-1}$ $E_{2}$ & 157.246062 & 36.3 & -4.67809 \\
		      & $1_{0}-1_{-1}$ $E_{2}$ & 157.270832 & 15.4 & -4.65644 \\
		      & $3_{0}-3_{-1}$ $E_{2}$ & 157.272338 & 27.1 & -4.66831 \\
		      & $2_{0}-2_{-1}$ $E_{2}$ & 157.276019 & 20.1 & -4.66116 \\
\enddata
\tablecomments{Taken from the Jet Propulsion Laboratory (JPL) catalog \citep{1998JQSRT..60..883P}.}
\end{deluxetable}

We carried out data reduction and imaging using the Common Astronomy Software Application \citep[CASA v 5.4.0;][]{2007ASPC..376..127M} on the pipeline-calibrated visibilities.
The data cubes were created with the CASA tclean task.
Uniform weighting was applied. 
The resulting angular resolution of $\sim 11\arcsec \times 8\arcsec$ corresponds to $\sim$ 0.016 pc $\times$ 0.011 pc ($\sim$ 3220 au $\times$ 2340 au) at the source distance \citep[293 pc;][]{2018ApJ...865...73O}.
The pixel size and image size are 2\farcs3 and 250 $\times$ 250 pixels.

Continuum ($\lambda = 2$ mm) images were created from the data cubes using the IMCONTSUB task.
The noise levels of the continuum images are 4 mJy beam$^{-1}$ in IRAS 03235 and IRAS 03245, and 1 mJy beam$^{-1}$ in IRAS 03271, respectively.

\section{Results and Analyses} \label{sec:resana}

\subsection{Results: Spatial distributions} \label{sec:res}

\begin{figure*}[!th]
\figurenum{1}
 \begin{center}
  \includegraphics[bb= 30 25 810 291, scale=0.6]{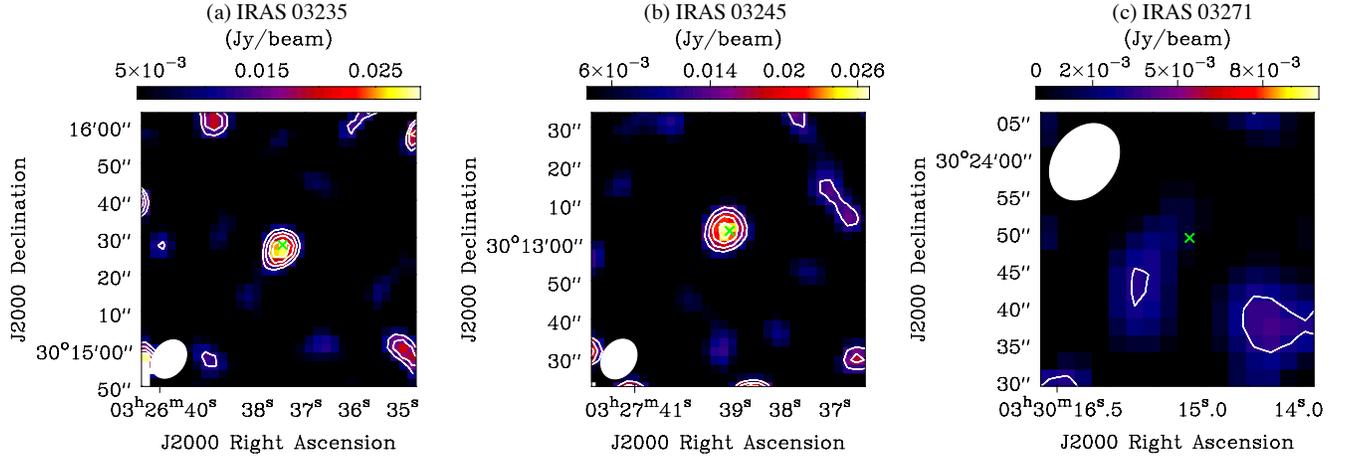}
 \end{center}
\caption{Continuum images toward the three IRAS sources. The contour levels are $3\sigma$, $4\sigma$, $5\sigma$, and $6\sigma$ of the rms noise levels. In panel (c), only the $3\sigma$ level contour appears. The noise levels are 4 mJy beam$^{-1}$ for panels (a) and (b), and 1 mJy beam$^{-1}$ for panel (c), respectively. The filled white ellipses indicate the angular resolution of approximately 11\farcs0 $\times$ 8\farcs6. The crosses indicate source positions taken from the Spitzer c2d program \citep{2009ApJ...692..973E}. \label{fig:cont}}
\end{figure*}

\begin{figure*}[!th]
\figurenum{2}
 \begin{center}
  \includegraphics[scale=0.7]{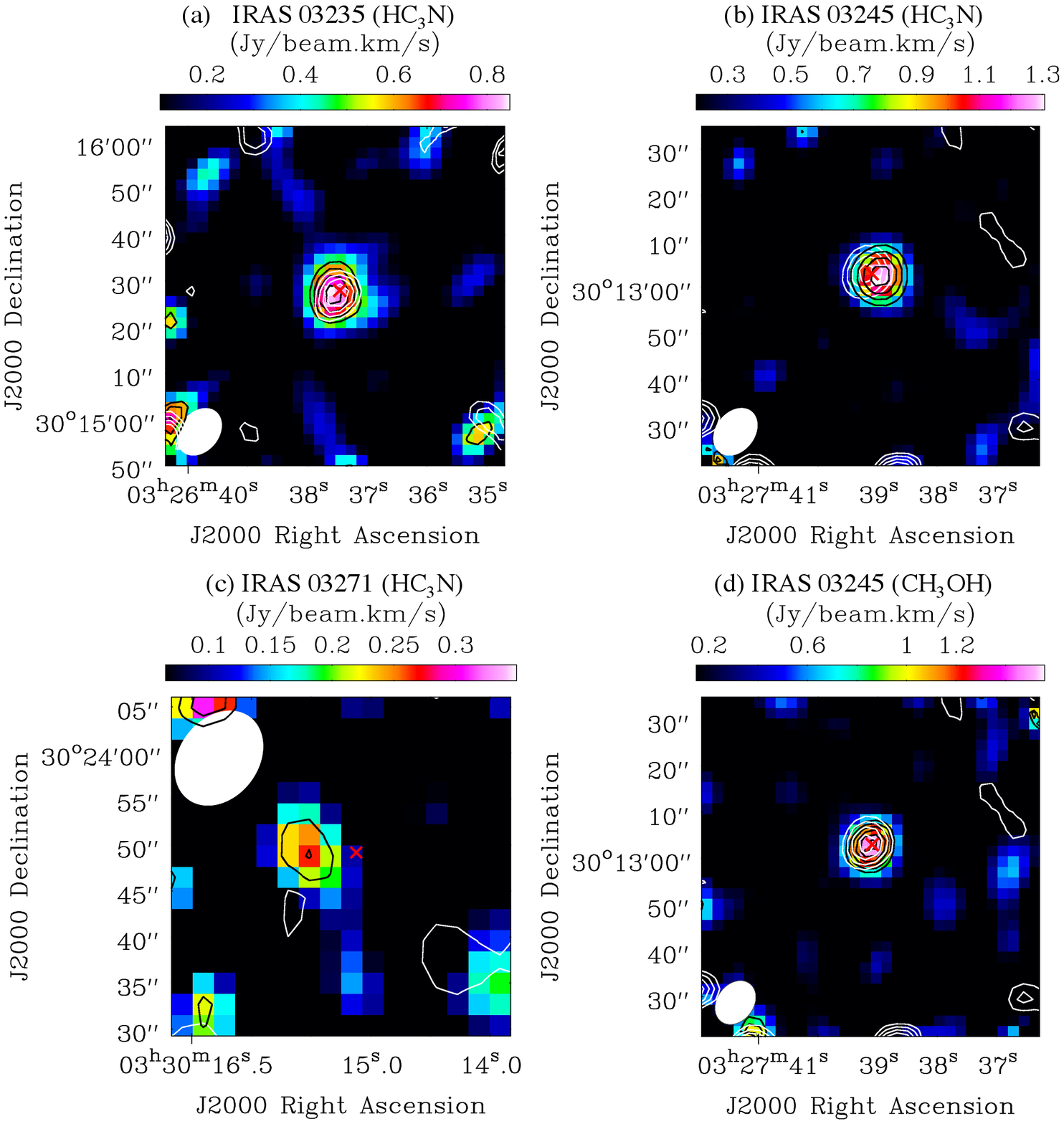}
 \end{center}
\caption{Moment 0 maps of (a)--(c) HC$_{3}$N and (d) CH$_{3}$OH. The black contour levels are $5\sigma$, $6\sigma$, $7\sigma$, and $8\sigma$ for panels (a) and (d), $3\sigma$, $4\sigma$, $5\sigma$, and $6\sigma$ for panels (b) and (c), respectively. In panel (c), only the $3\sigma$ and $4\sigma$ level contours appear. The noise levels are 0.1, 0.2, 0.065, and 0.15 Jy beam$^{-1}$ $\times$ km s$^{-1}$ for panels (a)--(d), respectively. The white contours show the distribution of dust continuum emissions, as in Figure \ref{fig:cont}. The filled white ellipses represent an angular resolution of approximately 11\farcs2 $\times$ 8\farcs4 for panels (a)--(c) and 10\farcs6 $\times$ 8\farcs2 for panel (d).  The crosses indicate source positions from the Spitzer c2d program \citep{2009ApJ...692..973E}. \label{fig:mom0}}
\end{figure*}

Figure \ref{fig:cont} shows the continuum images of the three target sources.
The green crosses indicate positions of infrared sources revealed by the Spitzer Core to Disk (c2d) Legacy program \citep{2003PASP..115..965E,2009ApJ...692..973E}. 
In IRAS 03235 and IRAS 03245, the dust continuum peaks are consistent with the Spitzer sources.
On the other hand, the detection of dust continuum emission is not significant in IRAS 03271.

The HC$_{3}$N ($J=16-15$) line has been detected in all of the sources, and panels (a)--(c) of Figure \ref{fig:mom0} show moment 0 maps of HC$_{3}$N toward the three sources.
Methanol (CH$_{3}$OH) has been detected only in IRAS 03245.
We used two lines ($1_{0}-1_{-1}$ and $3_{0}-3_{-1}$) for making the CH$_{3}$OH moment 0 map, because these lines are partially blended.
Panel (d) of Figure \ref{fig:mom0} shows its moment 0 map toward IRAS 03245.

Since all of the target sources have similar envelope masses (Table \ref{tab:source}), the detection/non-detection of CH$_{3}$OH does not likely depend on the presence of an envelope.
\citet{2016ApJ...819..140G} detected CH$_{3}$OH toward all of the three sources with the IRAM 30-m telescope with a beam size of $\sim 25$\arcsec.
In their observations, only two lines with low upper-state energies ($E_{\rm {up}}/k=6.97$ K and 12.54 K) were detected as weak lines in IRAS 03235 and IRAS 03271.
The noise levels of the ACA 7-m observations could be too high to detect weak lines of CH$_{3}$OH in IRAS 03235 and IRAS 03271.

We applied 2D gaussian fitting to the continuum and moment 0 maps. 
The fit results are summarized in Table \ref{tab:2Dgauss}.
In the case that beam-deconvolved sizes could not be obtained, we indicate the synthesized beam as their upper limits.

In IRAS 03235, the peak position of HC$_{3}$N is consistent with that of continuum image as well as the Spitzer source.
The size of the HC$_{3}$N emission ($\sim11\arcsec$, corresponding to $\sim 3223$ au) may be slightly larger than that of the continuum, as also seen in panel (a) of Figure \ref{fig:mom0}. 
The extent of the HC$_{3}$N emission ($\sim 3223$ au) is similar to the size of spatial distributions of carbon-chain species in L1527 in the Taurus region \citep{2010ApJ...722.1633S}.
Thus, HC$_{3}$N is likely to be efficiently formed in moderate temperature regions ($\sim 25-35$ K) by the warm carbon chain chemistry \citep[WCCC;][]{2013ChRv..113.8981S} mechanism.

In IRAS 03245, the peak positions of dust continuum and CH$_{3}$OH are consistent with the Spitzer point source. 
The spatial extent of HC$_{3}$N ($\approx 10\arcsec$, $\sim 2930$ au) is similar to that in IRAS 03235 ($\sim 3223$ au).
Again, the WCCC mechanism could contribute to formation of HC$_{3}$N. 
The emission region of HC$_{3}$N is larger than that of CH$_{3}$OH ($\sim 6 \arcsec$, corresponding to $\sim 1758$ au).
These features can be seen in panels (b) and (d) of Figure \ref{fig:mom0}.

The peak position of HC$_{3}$N emission is not consistent with the Spitzer point source, but is located $\sim 1470$ au to the east of it.
The origins of such discrepancies are not clear at the current angular resolution.
One possibility is that HC$_{3}$N emission is associated with the molecular outflow, because the direction of the offset of the HC$_{3}$N emission peak from the Spitzer source is similar to the direction of the molecular outflow \citep{2019ApJ...884..149H}.

\begin{deluxetable*}{cccccc}
\tablenum{4}
\tablecaption{Results of 2D gaussian fitting \label{tab:2Dgauss}}
\tablewidth{0pt}
\tablehead{
\colhead{Source} & \colhead{Species} & \colhead{Peak Position (J2000)} & \colhead{Size} &  \colhead{Position Angle} & \colhead{Peak Intensity\tablenotemark{a}} 
}
\startdata
IRAS 03235 & continuum & $03^{\rm {h}}26^{\rm {m}}$37\fs507 $\pm 0\fs013$, +30\degr15\arcmin26\farcs81 $\pm 0\farcs20$ & $< 11\farcs0 \times 8\farcs5$ & ... & $33.0 \pm 1.7$ \\
		    & HC$_{3}$N & $03^{\rm {h}}26^{\rm {m}}$37\fs498 $\pm0\fs025$, +30\degr15\arcmin27\farcs83 $\pm 0\farcs36$ & $11\farcs5 \pm 1\farcs5 \times 8\farcs6 \pm 2\farcs0$ & 39\degr & $0.839 \pm 0.047$ \\
IRAS 03245 & continuum & $03^{\rm {h}}27^{\rm {m}}$39\fs171 $\pm 0\fs016$, +30\degr13\arcmin03\farcs28 $\pm 0\farcs23$ & $< 11\farcs0 \pm \times 8\farcs6$ & ... & $30.1 \pm 1.7$ \\
		    & HC$_{3}$N & $03^{\rm {h}}27^{\rm {m}}$38\fs971 $\pm 0\fs018$, +30\degr13\arcmin02\farcs57 $\pm 0\farcs21$ & $9\farcs92 \pm 0\farcs89 \times 2\farcs83 \pm 2\farcs53$ &55\degr & $1.306 \pm 0.054$ \\
	        & CH$_{3}$OH & $03^{\rm {h}}27^{\rm {m}}$39\fs129 $\pm 0\fs010$, +30\degr13\arcmin02\farcs60 $\pm 0\farcs14$ & $5\farcs82 \pm 0\farcs78 \times 1\farcs25 \pm 1\farcs64$ & 49\degr & $1.659 \pm 0.051$ \\
IRAS 03271 & HC$_{3}$N & $03^{\rm {h}}30^{\rm {m}}$15\fs494 $\pm 0\fs025$, +30\degr23\arcmin49\farcs18 $\pm 0\farcs63$ & $< 11\farcs1 \times 8\farcs5$ & ... & $0.305 \pm 0.035 $ \\
\enddata
\tablenotetext{a}{Units for Peak Intensity are ``mJy beam$^{-1}$" and ``Jy beam$^{-1}$ $\times$ km s$^{-1}$" for continuum and the molecular emissions, respectively.}
\end{deluxetable*}

\subsection{Spectral analyses} \label{sec:ana}

\begin{figure*}[!th]
\figurenum{3}
 \begin{center}
  \includegraphics[bb = 0 25 540 570, scale=0.85]{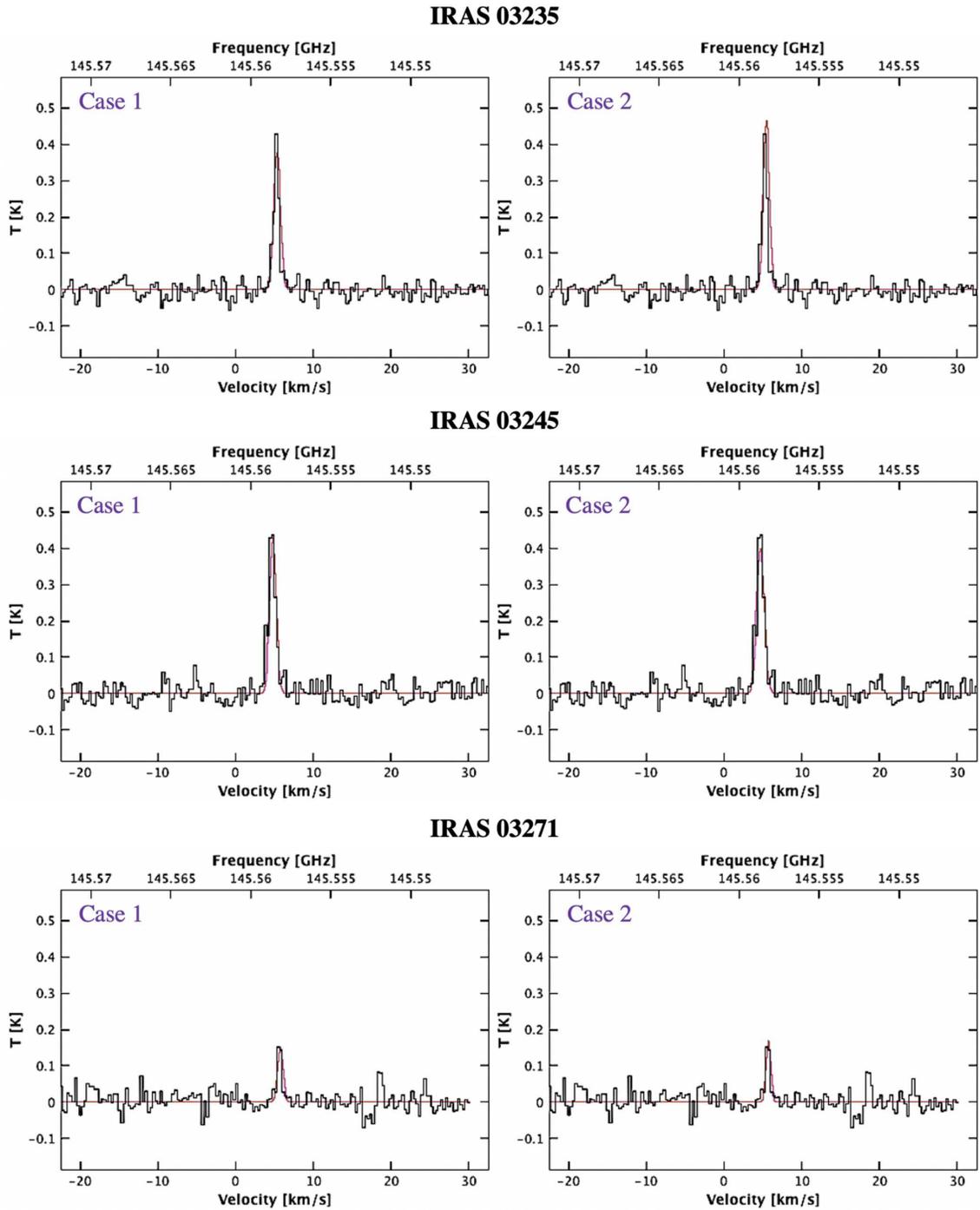}
 \end{center}
\caption{Spectra of HC$_{3}$N ($J=16-15$) toward the positions of peak emission of this molecule. The purple lines are modeled spectra of the best fitted results with CASSIS. \label{fig:HC3N}}
\end{figure*}

\begin{figure*}[!th]
\figurenum{4}
 \begin{center}
  \includegraphics[bb = 0 25 540 390, scale=0.85]{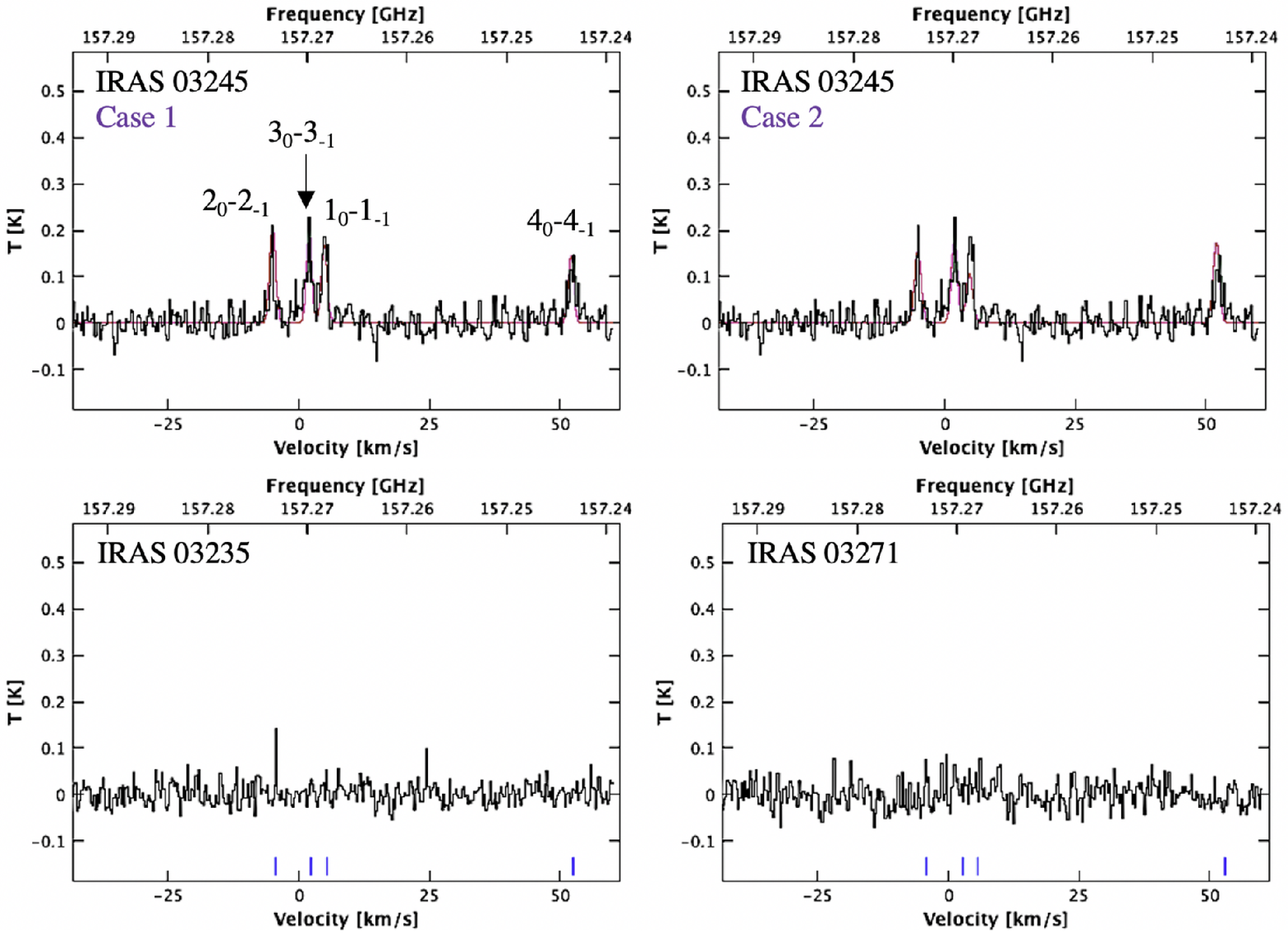}
 \end{center}
\caption{Spectra of CH$_{3}$OH ($4_{0}-4_{-1}$ $E_{2}$, $1_{0}-1_{-1}$ $E_{2}$, $3_{0}-3_{-1}$ $E_{2}$, and $2_{0}-2_{-1}$ $E_{2}$) at the HC$_{3}$N emission peaks. The purple lines are modeled spectra of the best fitted results with CASSIS. In panels of IRAS 03235 and IRAS 03271, the blue lines indicate the positions of the CH$_{3}$OH lines with the same $V_{\rm {LSR}}$ values as those of HC$_{3}$N, 5.4 km s$^{-1}$ and 5.7 km s$^{-1}$ for panels of IRAS 03235 and IRAS 03271, respectively. \label{fig:CH3OH}}
\end{figure*}

Figures \ref{fig:HC3N} and \ref{fig:CH3OH} show spectra of HC$_{3}$N and CH$_{3}$OH at the HC$_{3}$N emission peaks (Table \ref{tab:2Dgauss}) with beam sizes of $14\arcsec$, $12\arcsec$, and $9\arcsec$ in IRAS 03235, IRAS 03245, and IRAS 03271.
These beam sizes correspond to the size of the HC$_{3}$N emission in its moment 0 maps. 
In the IRAS 03235 and IRAS 03271 panels of Figure \ref{fig:CH3OH}, blue lines indicate velocity positions of 5.4 km s$^{-1}$ and 5.7 km s$^{-1}$, which are the $V_{\rm {LSR}}$ values of HC$_{3}$N in each source.

We analyzed spectra using the CASSIS software \citep{2011IAUS..280P.120C}.
In the analyses, we used the local thermodynamic equilibrium (LTE) model available in the CASSIS spectrum analyzer, assuming that the lines are optically thin.
Using the Markov chain Monte Carlo (MCMC) method and the LTE model, we derived the column densities ($N$) and excitation temperatures ($T_{\rm {ex}}$) of HC$_{3}$N and CH$_{3}$OH, treating $N$, $T_{\rm {ex}}$, line width (FWHM), and $V_{\rm {LSR}}$ as semi-free parameters within certain ranges.
In the MCMC method, the minimum and maximum values set for parameter in a component define the bounds in which the values are chosen randomly.
We consider the following two cases:
\begin{enumerate}
\item The range of the excitation temperature of both HC$_{3}$N and CH$_{3}$OH is 15--30 K; hereafter Case 1.
\item The excitation temperatures are comparable to dust temperatures (Table \ref{tab:source}); the excitation temperature ranges are set at 35--40 K in IRAS 03235 and IRAS 03245, and 35--50 K in IRAS 03271 (Case 2).
\end{enumerate}
We consider the first case because the WCCC mechanism \citep{2013ChRv..113.8981S} seems to work in our target sources as mentioned in Section \ref{sec:res}.
The assumed temperature range covers typical excitation temperatures of carbon-chain species in L1527 \citep{2019PASJ...71S..18Y}. 
We adopt the second case because the observed lines may come from inner dense cores compared with previous single-dish observations \citep{2016ApJ...819..140G, 2017ApJ...841..120B, 2019PASJ...71S..18Y}, where contributions from outer envelopes may be significant due to low upper-state-energy lines, and the excitation temperature may be underestimated in our data.

The obtained parameters for each case are summarized in Table \ref{tab:CASSIS}.
The modeled spectra using the best fitted values are overlaid as purple lines in Figures \ref{fig:HC3N} and \ref{fig:CH3OH}.
The line widths (FWHM) of HC$_{3}$N in our target sources ($\sim 1$ km s$^{-1}$) are larger than the typical value in starless cores \citep[$\sim 0.5$ km s$^{-1}$;][]{2004PASJ...56...69K} but similar to that in the L1527 low-mass WCCC source \citep{2019PASJ...71S..18Y}.
Thus, the detected HC$_{3}$N line seems to mainly trace lukewarm regions and not come from outer cold regions.
In IRAS 03245, the $V_{\rm {LSR}}$ value of HC$_{3}$N is consistent with that of CH$_{3}$OH.

The obtained excitation temperatures of HC$_{3}$N in all of the three sources in Case 1 are $16-18$ K, which are comparable with those in L1527 \citep{2019PASJ...71S..18Y}.
The derived column densities of HC$_{3}$N change by only a factor of $\sim 3-4$ between Cases 1 and 2.
The column densities of HC$_{3}$N in Case 1 are slightly higher than the previous results obtained with the IRAM 30-m telescope \citep[the telescope beam size is $\sim$ 25\arcsec;][]{2017ApJ...841..120B} by a factor of $\approx 1.3-2$.
\citet{2017ApJ...841..120B} derived the HC$_{3}$N column densities by the rotational diagram method using the two lines with almost similar upper-state energies or using only one line with assumed fixed rotational temperatures.
The effect of the beam dilution could also result in the lower column densities in the single-dish observations \citep{2017ApJ...841..120B}.

The CH$_{3}$OH column densities in IRAS 03245 for Cases 1 and 2 are higher than that derived by \citet{2016ApJ...833..125G} by a factor of 1.8--2.8.
\citet{2016ApJ...833..125G} derived the CH$_{3}$OH column density to be ($1.5 \pm 0.3$)$\times 10^{13}$ cm$^{-2}$ from observations with the IRAM 30-m telescope.
Again, this difference in column density could be caused by the beam dilution in the single-dish observations.

\floattable
\rotate
\begin{deluxetable}{ccccccccccc}
\tablenum{5}
\tablecaption{Obtained parameters from the spectral analyses \label{tab:CASSIS}}
\tablewidth{0pt}
\tablehead{
\colhead{Source} & \colhead{Species} & \multicolumn{4}{c}{Case 1} & \colhead{} & \multicolumn{4}{c}{Case 2} \\
\cline{3-6} \cline{8-11}
\colhead{} & \colhead{} & \colhead{$N$} & \colhead{$T_{\rm {ex}}$} & \colhead{FWHM} & \colhead{$V_{\rm {LSR}}$} & \colhead{} & \colhead{$N$} & \colhead{$T_{\rm {ex}}$} & \colhead{FWHM} & \colhead{$V_{\rm {LSR}}$} \\
\colhead{} & \colhead{} & \colhead{(cm$^{-2}$)} & \colhead{(K)} & \colhead{(km s$^{-1}$)} & \colhead{(km s$^{-1}$)} & \colhead{} & \colhead{(cm$^{-2}$)} & \colhead{(K)} & \colhead{(km s$^{-1}$)} & \colhead{(km s$^{-1}$)}
}
\startdata
IRAS 03235 & HC$_{3}$N & ($4.9 \pm 0.4$)$\times 10^{12}$ & $18.0 \pm 0.9$ & $1.06 \pm  0.05$ & $5.38 \pm 0.02$ & & ($1.3 \pm 0.7$)$\times10^{12}$ & $37.3 \pm 0.2$ & $0.83 \pm 0.06$ & $5.43 \pm 0.02$ \\
IRAS 03245 & HC$_{3}$N & ($7.2 \pm 0.2$)$\times 10^{12}$ & $16.3 \pm 0.4$ & $0.989 \pm 0.006$ & $4.821 \pm 0.017$ & & ($2.7 \pm 1.1$)$\times 10^{12}$ & $37.3 \pm 0.3$ & $1.05 \pm 0.07$ & $4.72 \pm 0.05$ \\ 
		  & CH$_{3}$OH & ($2.7 \pm 0.4$)$\times 10^{13}$ & $17.5 \pm 0.4$ & $1.30 \pm 0.13$ & $4.86 \pm 0.05$ & & ($4.3 \pm 0.5$)$\times 10^{13}$ & $37.4 \pm 1.5$ & $1.50 \pm 0.17$ & $4.74 \pm 0.04$ \\
IRAS 03271 & HC$_{3}$N & ($2.2 \pm 0.3$)$\times 10^{12}$ & $16.2 \pm 0.8$ & $0.82 \pm 0.11$ & $5.745 \pm 0.012$ & & ($5.3 \pm 1.6$)$\times 10^{11}$ & $46.6 \pm 1.6$ & $0.71 \pm 0.16$ & $5.72 \pm 0.04$ \\	  
\enddata
\tablecomments{The errors represent the standard deviation.}
\end{deluxetable}

\section{Discussion} \label{sec:dis}

We now compare the derived CH$_{3}$OH/HC$_{3}$N column-density ratios with those derived by the chemical network simulations.
We describe our chemical network simulations in Subsection \ref{sec:dis1}, and compare observations and simulations in Subsection \ref{sec:dis2}.

\subsection{Chemical network simulations} \label{sec:dis1}

We used the chemical network code Nautilus \citep{2016MNRAS.459.3756R} and a hot-core model with a warm-up stage. 
The physical evolution is the same as that in \citet{2019ApJ...881...57T}.
The initial gas density is $n_{\rm {H}}=10^{4}$ cm$^{-3}$ and increases to $10^{7}$ cm$^{-3}$ during the freefall collapse.
The initial temperature is 10 K and rises to 200 K during the warm-up period.
In this paper, we investigated low-mass YSOs, and thus we utilized the model with a slow warm-up period ($1 \times 10^{6}$ yr).
The dust temperature is assumed to be equal to the gas temperature.

We changed the cosmic ray ionization rate ($\zeta$) and the initial C/O ratio in order to investigate the dependence of the CH$_{3}$OH/HC$_{3}$N ratio on these parameters.
Table \ref{tab:model} summarizes the models used in this paper.
Models No.\,1--No.\,3 are the same as those presented in \citet{2019ApJ...881...57T}.
The C/O elemental ratio could change the CH$_{3}$OH and HC$_{3}$N abundances.
We included the high cosmic ray ionization rates because this rate ($\zeta = 4.0 \times 10^{-14}$ s$^{-1}$) can reproduce the observed abundances of carbon-chain species in intermediate-mass protoclusters \citep{2017A&A...605A..57F,2018ApJ...859..136F} and around some massive young stellar objects (MYSOs) \citep{2019ApJ...881...57T, 2020arXiv201212993T}.

\begin{deluxetable}{ccc}
\tablenum{6}
\tablecaption{Summary of models \label{tab:model}}
\tablewidth{0pt}
\tablehead{
\colhead{Model} & \colhead{$\zeta$ (s$^{-1}$)\tablenotemark{a}} & \colhead{C/O ratio}
}
\startdata
No.\,1 & $1.3 \times 10^{-17}$ & 0.4\tablenotemark{b} \\
No.\,2 & $3.0 \times 10^{-16}$ & 0.4\tablenotemark{b} \\
No.\,3 & $4.0 \times 10^{-14}$ & 0.4\tablenotemark{b} \\
No.\,4 & $1.3 \times 10^{-17}$ & 0.5\tablenotemark{c} \\
No.\,5 & $1.3 \times 10^{-17}$ & 1.2\tablenotemark{d} \\
\enddata
\tablenotetext{a}{Cosmic ray ionization rates used in the model}
\tablenotetext{b}{The C/O ratio of 0.4 corresponds to C$^{+}$ = $7.3 \times 10^{-5}$ and O = $1.76 \times 10^{-4}$, and the rest of the elemental abundances remain the same as described in \citet{2019ApJ...881...57T}.}
\tablenotetext{c}{Corresponding to C$^{+}$ = $1.7 \times 10^{-4}$ and O = $3.3 \times 10^{-4}$.}
\tablenotetext{d}{Corresponding to C$^{+}$ = $1.7\times 10^{-4}$ and O = $1.4 \times 10^{-4}$.}
\end{deluxetable}

\begin{figure*}[!th]
\figurenum{5}
 \begin{center}
  \includegraphics[bb = 0 20 362 498, scale=0.8]{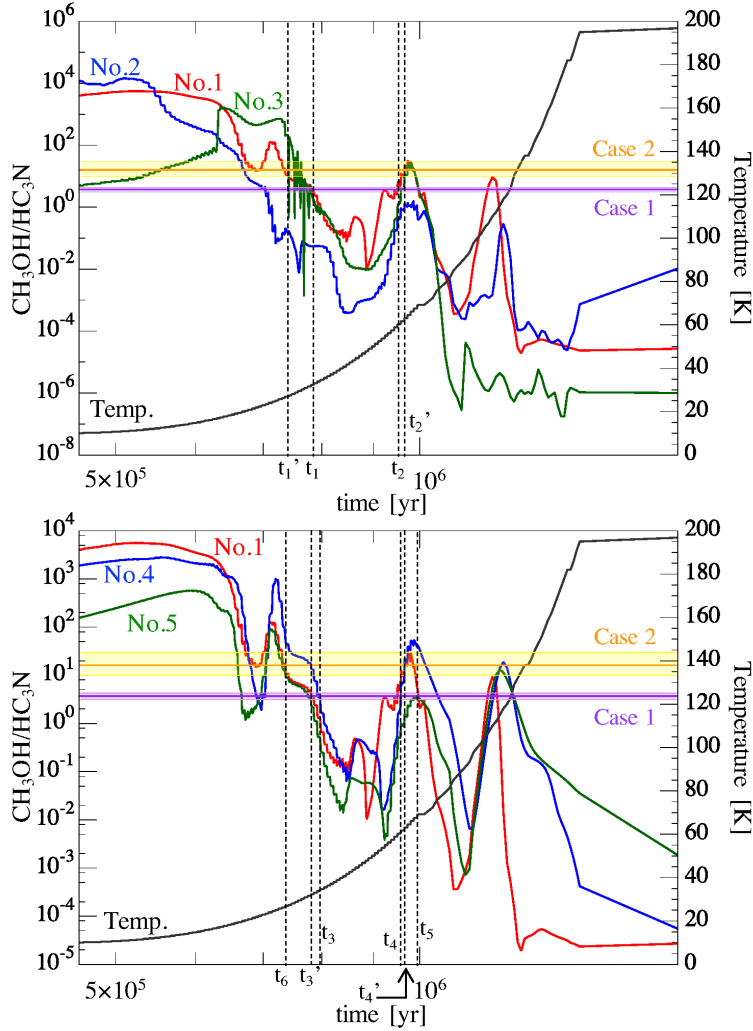}
 \end{center}
\caption{Comparisons of the CH$_{3}$OH/HC$_{3}$N ratios between observations and chemical network simulations (Models No.\,1--No.\,5, see Table \ref{tab:model}). The purple and orange horizontal lines indicate the observed values in IRAS 03245 for Case 1 and Case 2, respectively. The black curves indicate gas (and dust) temperature evolution from the model. The symbols of t$_{n}$ and t$_{n}$', ($n=1-6$) associated with dashed vertical lines represent the ages when the observed CH$_{3}$OH/HC$_{3}$N ratios agree with the modeled values. The details are described in the main text and values are summarized in Table \ref{tab:t}. \label{fig:model}}
\end{figure*}

Figure \ref{fig:model} shows the time dependence of the CH$_{3}$OH/HC$_{3}$N ratio during the warm-up period, taking the evolutionary stages of our target sources into consideration.
As a general trend of these models, the CH$_{3}$OH/HC$_{3}$N ratio decreases after $\sim 7\times10^{5}$ yr (23 K $<T<50$ K), slightly increases during $\sim 9\times10^{5}-10^{6}$ yr (50 K $<T<70$ K), and then drops again.
Such complicated features are caused by combinations of the temporal or thermal dependence of the abundance of CH$_{3}$OH and HC$_{3}$N during the warm-up stage.

The detail formation pathways of HC$_{3}$N during the warm-up stage are discussed in detail in \citet{2019ApJ...881...57T} and we include only a short description here.
Methane (CH$_{4}$) desorbs from ice mantles at temperatures above 25 K, and rapidly starts forming carbon-chain species in the gas phase, via the WCCC mechanism \citep{2008ApJ...681.1385H}.
This explains the first decrease in the CH$_{3}$OH/HC$_{3}$N ratio, because there is no efficient formation pathway for the gas-phase CH$_{3}$OH at that time ($T \approx 25-50$ K, $t \approx 7.2 \times 10^{5} - 9 \times 10^{5}$ yr). 
After HC$_{3}$N is formed in the gas phase, it is soon adsorbed onto dust grains and accumulated in ice mantles, leading to a decrease in the gas-phase HC$_{3}$N abundance.

Methanol (CH$_{3}$OH) is mainly formed by the electron recombination reaction of CH$_{3}$OCH$_{4}^{+}$ ($t \approx 7\times10^{5}-10^{6}$ yr) in the gas phase, before the temperature reaches its sublimation temperature.
The formation of the CH$_{3}$OCH$_{4}^{+}$ ion becomes enhanced by the reaction between HCO$^{+}$ and CH$_{3}$OCH$_{3}$, for which the formation increases in rate by the associative gas-phase reaction between CH$_{3}$O and CH$_{3}$ from $t \approx 7 \times 10^{5}$ yr.

After the temperature reaches at 87 K ($t=1.09\times10^{6}$ yr), CH$_{3}$OH desorbs from dust grains and its gas phase abundance increases sharply, while HC$_{3}$N starts desorbing from dust grains at $t\simeq1.1\times10^{6}$ yr ($T\simeq90$ K).

\subsection{Comparison of the CH$_{3}$OH/HC$_{3}$N abundance ratio between observations and models} \label{sec:dis2}

\begin{deluxetable}{cccc}
\tablenum{7}
\tablecaption{Summary of ages at which models agree with the observed CH$_{3}$OH/HC$_{3}$N ratio and corresponding temperatures \label{tab:t}}
\tablewidth{0pt}
\tablehead{
\colhead{Label} & \colhead{Age} & \colhead{Temperature} & \colhead{pair of} \\
\colhead{} & \colhead{(yr)} & \colhead{(K)} & \colhead{Model and observation}
}
\startdata
t$_{1}$ & $7.8 \times 10^{5}$ & 32 & No.\,1, 3, 5 \& Case 1 \\
t$_{2}$ & (9.2--9.3)$\times10^{5}$ & 55 & No.\,1, 3 \& Case 1 \\
t$_{1}$' & (7.3--7.5)$\times 10^{5}$ & 26--28 & No.\,1, 3 \& Case 2 \\
t$_{2}$' & (9.6--9.7)$\times 10^{5}$ & 62--65 & No.\,1, 3 \& Case 2 \\
t$_{3}$ & $7.9 \times 10^{5}$ & 34 & No.\,4 \& Case 1 \\
t$_{4}$ &  $9.5 \times 10^{5}$ & 60 & No.\,4 \& Case 1 \\
t$_{3}$' & (7.7--7.8)$\times 10^{5}$ & 31--33 & No.\,4 \& Case 2 \\
t$_{4}$' & $9.7 \times 10^{5}$ & 63 & No.\,4 \& Case 2 \\
t$_{5}$ & (9.8--10)$\times 10^{5}$ & 66-69 & No.\,5 \& Case 1 \\
t$_{6}$ & $7.3 \times 10^{5}$ & 26 & No.\,5 \& Case 2
\enddata
\end{deluxetable}

The observed CH$_{3}$OH/HC$_{3}$N ratios in IRAS 03245 are derived to be $3.7 \pm 0.6$ and $16^{+14}_{-6}$ for Case 1 and Case 2, respectively, using the results summarized in Table \ref{tab:CASSIS}.
The upper panel of Figure \ref{fig:model} shows the time dependence of the CH$_{3}$OH/HC$_{3}$N ratio for three models with the different cosmic ray ionization rates (Models No.\,1--No.\,3), while the lower panel shows three models with the different C/O ratios (Models No.\,1, No.\,4, and No.\,5).
The observed CH$_{3}$OH/HC$_{3}$N ratios are indicated as horizontal lines including error bars; purple and orange lines indicate the results of Case 1 and Case 2, respectively.
The temperature scale arises from only the model.
We now focus on comparison with observations, starting with ages after the temperature reaches 25 K, at which the WCCC mechanism starts ($t \geq 7.2 \times 10^{5}$ yr).

We first compare the observed ratio in IRAS 03245 to Model No.\,1.
The observed ratio for Case 1 ($3.7 \pm 0.6$) agrees with Model No.\,1 at $t\approx7.8 \times 10^{5}$ yr (labeled as t$_{1}$), at which the temperature is $\simeq 32$ K, as shown in the upper panel of Figure \ref{fig:model}.
This dust temperature agrees with the dust temperature derived in IRAS 03245 \citep[$37 \pm 2$ K;][see also Table \ref{tab:source} in this paper]{2009AA...493...89E}.
Around $t\approx$($9.2-9.3$)$\times10^{5}$ yr (t$_{2}$, $T\approx55$ K), the observed ratio in IRAS 03245 again agrees with Model No.\,1, but the temperature is slightly higher than the dust temperature in IRAS 03245.

The observed ratio in IRAS 03245 with Case 2 ($16^{+14}_{-6}$) is consistent with Model No.\,1 around ($7.3-7.5$)$\times 10^{5}$ yr (t$_{1}$', $T\simeq26-28$ K).
The dust temperature of 26--28 K is slightly lower than the observed dust temperature in this source.
Since we assumed that the higher temperature conditions in Case 2 compared with Case 1 and the excitation temperatures of HC$_{3}$N and CH$_{3}$OH are derived to be around 37 K in Case 2 (Table \ref{tab:CASSIS}), such a lower temperature (26--28 K) compared to Case 1 ($\simeq 32$ K) is contradictory.
Besides, the CH$_{3}$OH/HC$_{3}$N ratio with Case 2 agrees with Model No.\,1 around ($9.6-9.7$)$\times 10^{5}$ yr (t$_{2}$', 62--65 K).
However, the temperature at this age is higher than the observed dust temperature in this source.

In the similar manner, we investigate the ages and temperatures when the observed CH$_{3}$OH/HC$_{3}$N ratios match those derived by other models (No.\,2 -- No.\,5), and the results are summarized in Table \ref{tab:t}.
In Model No.\,2, the modeled CH$_{3}$OH/HC$_{3}$N ratio is lower than the observed ratios in IRAS 03245 for Cases 1 and 2 after $t \geq 7.2 \times 10^{5}$ yr, so that Model No.\,2 cannot reproduce the observed ratio in this source.

Taking all of the results summarized in Table \ref{tab:t} into consideration, the observed CH$_{3}$OH/HC$_{3}$N column-density ratio and dust temperature in IRAS 03245 can be best reproduced simultaneously by our models at $t \approx$ (7.7--7.9)$\times 10^{5}$ yr ($T \approx 31-34$ K).
This temperature range agrees with the WCCC mechanism \citep{2008ApJ...681.1385H}.
The other temperature ranges ($\sim 25-28$ K and $\sim 60-70$ K) are not consistent with the observed dust temperatures, and hence these ages and temperatures are unlikely.

In IRAS 03235 and IRAS 03271, we could not detect CH$_{3}$OH, and the CH$_{3}$OH/HC$_{3}$N ratio should be lower than that in IRAS 03245 ($< 3.7$).
In that case, the CH$_{3}$OH/HC$_{3}$N ratios toward these two low-mass YSOs agree with Model No.\,1 around t$_{1}$($7.8 \times 10^{5}$ yr) $<t<$ t$_{2}$ ($9.3\times 10^{5}$ yr), corresponding to 32 K $< T < 55$ K.
Although the dust temperature was not derived in IRAS 03235, it is expected that it is not so far from those in IRAS 03245 ($37 \pm 2$ K) and IRAS 03271 ($45 \pm 2$ K) because of their similar ages and physical properties (see Table \ref{tab:source}).
If this is the case, the suggested temperature range agrees with the dust temperatures in IRAS 03235 and IRAS 03271.
In the model calculations, the CH$_{3}$OH/HC$_{3}$N ratios become lower than the observed ratios in IRAS 03245 after $\approx 10^{6}$ yr.
However, the dust temperature becomes higher than 70 K after $\approx 10^{6}$ yr.
These dust temperatures above 70 K are not consistent with the observed dust temperatures in our target sources.
Thus, we omit a possibility of ages after $\approx 10^{6}$ yr.

Since the dust temperatures in the observed sources do not reach the sublimation temperature of CH$_{3}$OH, we conclude that the non-thermal desorption mechanisms or gas-phase formation reactions leading to CH$_{3}$OH are the primary contributors to gas-phase CH$_{3}$OH formation around low-mass Class 0 and Class I YSOs at the time/temperature regime studied here.
These mechanisms seem to be especially efficient in IRAS 03245, among the observed three sources.
In the chemical network simulation, gas-phase ion-molecule reactions contribute to the formation of gas-phase CH$_{3}$OH, via the electron recombination reaction of CH$_{3}$OCH$_{4}^{+}$.
These hypotheses for gas-phase CH$_{3}$OH formation, in particular those involving non-thermal desorption such as photo-desorption, are supported by the fact that the bolometric luminosity in IRAS 03245 is higher than the other two sources.

\section{Conclusions} \label{sec:con}

We have analyzed ALMA Cycle 5 data in Band 4 toward three low-mass YSOs: IRAS 03235+3004 (IRAS 03235), IRAS 03245+3002 (IRAS 03245), and IRAS 03271+3013 (IRAS 03271), in the Perseus region.
The HC$_{3}$N ($J=16-15$) line was detected toward all sources, while the CH$_{3}$OH lines were detected only toward IRAS 03245.
The detection/non-detection of CH$_{3}$OH is independent of the mass of envelopes.
These results may imply the chemical diversity around low-mass YSOs.

In IRAS 03235 and IRAS 03245, both of which are Class 0 sources, the continuum peaks correspond to the Spitzer sources.
The spatial distributions of HC$_{3}$N are extended relative to those of the continuum emission in these two sources.
On the other hand, the spatial distribution of CH$_{3}$OH matches the continuum emission in IRAS 03245.
In IRAS 03271, the spatial distribution of the continuum emission shows an elongated feature and its peak does not correspond to the Spitzer source.
The spatial distribution of HC$_{3}$N differs from that of continuum emission and its peak does not match with the Spitzer source.
Although we could not understand the discrepancy between the spatial distributions of HC$_{3}$N and that of continuum with the current angular resolution, HC$_{3}$N may be originated from the molecular outflow. 

We derived the column densities and excitation temperatures of HC$_{3}$N and CH$_{3}$OH with the CASSIS software.
The CH$_{3}$OH/HC$_{3}$N ratio in IRAS 03245 was derived to be $3.7 \pm 0.6$ and $16^{+14}_{-6}$ on the assumptions that the excitation temperature is close to that of L1527, and the excitation temperature is close to the dust temperatures, respectively.
We compared the observed CH$_{3}$OH/HC$_{3}$N ratios to those derived by the chemical network simulations.
The observed CH$_{3}$OH/HC$_{3}$N ratios in IRAS 03245 are reproduced at ages when the dust temperature lie around $\approx 30-35$ K, which agrees with the dust temperature in IRAS 03245.
Regarding the other two sources, where CH$_{3}$OH has not been detected, the CH$_{3}$OH/HC$_{3}$N ratios, which should be lower than that in IRAS 03245, can be reproduced when the temperatures lie between 32 K $< T < 55$ K, which agrees with the dust temperatures in these low-mass YSOs.

At the times when the observed CH$_{3}$OH/HC$_{3}$N ratios and the dust temperatures are reproduced, HC$_{3}$N is efficiently formed by the warm carbon chain chemistry (WCCC) mechanism.
The non-thermal desorption or sublimation or the gas-phase reactions contribute to the formation of gas-phase CH$_{3}$OH during the times.  
The higher bolometric luminosity in IRAS 03245 seems to support these conclusions, especially the importance of non-thermal desorption mechanisms.

\acknowledgments
This paper makes use of the following ALMA data: ADS/JAO.ALMA\#2017.1.00955.S.
ALMA is a partnership of ESO (representing its member states), NSF (USA) and NINS (Japan), together with NRC (Canada), MOST and ASIAA (Taiwan), and KASI (Republic of Korea), in cooperation with the Republic of Chile. 
The Joint ALMA Observatory is operated by ESO, AUI/NRAO and NAOJ. 
Based on analyses carried out with the CASSIS software and JPL spectroscopic database. 
CASSIS has been developed by IRAP-UPS/CNRS (http://cassis.irap.omp.eu).
This work was supported by JSPS KAKENHI grant No. JP20K14523. 
This research was carried out in part at the Jet Propulsion Laboratory, which is operated for NASA by the California Institute of Technology.
E.H. thanks the National Science Foundation for support through grant AST-1906489.
We thank the anonymous referee who gave us valuable comments, which helped us improve the quality of the paper.

%

\vspace{5mm}
\facilities{Atacama Large Millimeter/submillimeter Array (ALMA)}


\software{Common Astronomy Software Applications package \citep[CASA;][]{2007ASPC..376..127M}, CASSIS \citep{2011IAUS..280P.120C}, Nautilus \citep{2016MNRAS.459.3756R}}







\end{document}